\documentclass[prb,twocolumn,preprintnumbers,amsmath,amssymb,floatfix,citeautoscript,nofootinbib]{revtex4}

\begin{document}

\title{Phase transition on the edge of the $\nu=\frac{5}{2}$ 
Pfaffian and
anti-Pfaffian quantum Hall state}% Force line breaks with \\

\author{B.J. Overbosch}
\author{Xiao-Gang Wen}%
\affiliation{%
Physics Department, Massachusetts Institute of Technology, Cambridge,
MA 02139, USA}
\date{April 13, 2008}% It is always \today, today,

\begin{abstract}
Starting from the edge reconstructed Pfaffian state and the anti-Pfaffian
state for the filling fraction $\nu=\frac{5}{2}$ fractional quantum Hall (FQH)
state with the filled Landau levels included, we find that interactions
between counterpropagating edge modes can induce phase transitions on the
edge. In the new `Majorana-gapped' phases, a pair of counter propagating
neutral Majorana modes becomes gapped. The quasiparticle tunneling exponent
changes from $g=1/4$ to $g=1/2$ for the  edge reconstructed Pfaffian state,
and changes from $g=1/2$ to $g=0.55-0.75$ for the anti-Pfaffian state,
in the new Majorana-gapped phases.
The new phases are candidate states for the observed $\nu=\frac{5}{2}$
state.
Furthermore, Majorana-gapped phases provide examples that non-trivial quantum
phase transitions can happen on the edge of a FQH state without any change in
bulk topological order.

\end{abstract}

\maketitle

%%%%%%%%%%%%%%%%%%%%%%%%%%%%%%%%%%%%%%%%%
%%%%%%%%%%%%%%%%%%%%%%%%%%%%%%%%%%%%%%%%%

\section{Introduction}

With the prospect of non-abelian anyons
and the potential
application of topologically protected qubits\cite{WNtop,K032,NSSFD0707.1889} there has been
a renewed interest in exotic fractional quantum Hall (FQH) states, especially
the experimentally easiest accessible state at filling fraction
$\nu=\frac{5}{2}$.\cite{WES8776,PYXetal0109} In 1991, two different
non-Abelian FQH states were proposed both at filling fraction $\nu=\frac{1}{2}$ (or
$\nu=\frac{5}{2}$).\cite{Wnab,MR9162} This raised an very interesting possibility that
the observed $\nu=\frac{5}{2}$ state may be a non-Abelian state.  One of the
proposed states, the Moore-Read Pfaffian wavefunction,\cite{MR9162} has
received a lot of attention recently, both in experiment and in theory.

However, it is currently still unclear what the true nature of the
$\nu=\frac{5}{2}$ bulk state is; many candidate states, or trial
wavefunctions, exist, some of which predict non-abelian statistics, others
predict less exotic abelian
statistics.\cite{H8375,MR9162,Wnab,Wctpt,LHR0706,LRNF0707}
We briefly review these candidates.

Experimentally the bulk state can be probed through transport measurements on
the edge, for instance through tunneling between FQH edges induced by a
quantum point contact
(QPC).\cite{Wedgetun,Wedgerev,CFK9731,SFN0502,FK0603}
The different candidate states make specific predictions for the charge $e^*$
and exponent $g$ of the tunneling quasiparticles, and these can in principal
be measured in experiment. In the regime of weak tunneling at sharp edges,
chiral Luttinger liquid ($\chi$LL) theory\cite{Wedgerev} describes the form of
e.g. the differential tunneling conductance and the tunneling current noise as
a function of temperature and applied bias, with $e^*$ and $g$ as
continuously-varying fitting parameters.

 Very recently two experimental groups reported results on tunneling
across a QPC in the $\nu=\frac{5}{2}$ state; {Ref.~}
\onlinecite{RMMKPW0803.3530} focusses on scaling of the differential
conductance, {Ref.~} \onlinecite{DHUSM0802.0930} measures
shot-noise. The results are most consistent with a quasiparticle charge $e^*=1/4$ and
exponent $g=1/2$.

In this paper we introduce two other candidates for the $\nu=\frac{5}{2}$ FQH
\emph{edge}. First, starting from the existing so-called `anti-Pfaffian' bulk
state,\cite{LHR0706,LRNF0707} we consider interactions between the counterpropagating
edge modes.  As we change the interaction strength, we find that there is a
transition to a new phase on the edge, with different values for $e^*$ and
$g$. 
Note that this is really
a phase transition on the edge, since the bulk state does not change. By
appropriately tuning the edge interactions one should be able to observe this
quantum phase transition through e.g. a change in $e^*$ and $g$.

We call this new phase the `Majorana-gapped' phase, as the anti-Pfaffian
Majorana mode becomes gapped.  The Majorana-gapped phase has 2 and 1/2
right-moving edge branches and 1 left-moving edge branches, while the standard
edge phase for the anti-Pfaffian state has 3 right-moving edge branches and 1
and 1/2 left-moving edge branches.  During the transition from the standard
edge phase to the Majorana-gapped phase, half a left-moving edge branch (a
Majorana fermion mode) pairs up with half a right-moving edge branch
and this opens up
an energy gap.

Second, we start with the edge reconstructed\cite{CW9427} Pfaffian
bulk state, which has
$2$ and $1/2$ branches of right movers and $1$ branch of left movers.  Such an
edge can also undergo a phase transition into a Majorana-gapped phase which has $2$
branches of right movers and $1/2$ branch of left movers.  The values of
$e^*$ and $g$ can be changed by the phase transition.%
\footnote{Here, with a change of $e^*$ we mean that another
quasiparticle with a different charge becomes the most dominant
quasiparticle which is observed in experiments. The phase transition
does not change the fixed charge $e^*$ of a given quasiparticle; it can change
the exponent $g$ for all quasiparticles.}

The above result is for the clean edge.  In the presence of impurities, the
picture is 
%quite 
different.  In that case, as we change the interaction strength between
different edge branches beyond a threshold, a right-moving Majorana fermion
mode pairs up with a left-moving Majorana fermion mode and they become
localized.  If we assume that the localized modes do not contribute to
tunneling between the edges, then we can treat those localized modes as if
they are gapped.  Under this assumption, the clean edge and dirty edge behave
similarly.

We like to stress that in order to have the gapping or the localization phase
transition, it is necessary to include the supposedly completely filled lowest
Landau level in the framework, or to include the additional edge branches from
the edge reconstruction. The new phases require that the different edge modes
 have substantial interactions with each other.

Numerical simulations for small-size closed
systems \cite{M9805,RH0085}, which by construction ignore edge effects, 
suggested that the Moore-Read Pfaffian trial wavefunction is the most likely
candidate for the actual $\nu=\frac{5}{2}$ FQH bulk state.
To compare
with actual experiments, which obviously have an edge, it is necessary, as our
examples in this paper shows, to include the edge-aspect as well.
This is emphasized as well in a very recent numerical study,
{Ref.~} \onlinecite{WHRY0816}, which considers a disc-geometry with an
edge, and a varying confining potential; for a sharp edge the Pfaffian
is found to be favored, but for a smooth edge the groundstate is a
different state which bears the marks of some form of edge reconstruction.

The `Majorana-gapping' transition is a quantum phase transition on the edge only
which does not affect the bulk state.  Such a kind of edge-only quantum phase
transition has been studied in {Ref.~} \onlinecite{KCW9963}.  Here we find a new
type of edge-only quantum phase transition where we lose (gain) a fractional
branch of right-movers (right-movers) through the transition.

Our paper is organized as follows. We review the different candidates
for the $\nu=\frac{5}{2}$ state in
Sec.~\ref{sec:candidatelist}. Section~\ref{sec:gappedantipf} is the
core of our paper. Here we show that for the anti-Pfaffian state there
exists an operator which for certain density-density
interactions becomes relevant and can drive a phase transition. In the
new phase a pair of counterpropagating Majorana modes becomes
gapped. For the new phase we determine the quasiparticle spectrum and
which of these quasiparticles is the most relevant.
In Sec.~\ref{sec:gappedpf} we apply the same formalism to the
edge reconstructed Pfaffian state. 
We discuss and summarize our results in Sec.~\ref{sec:discussion}.

%%%%%%%%%%%%%%%%%%%%%%%%%%%%%%%%%%%%%%%%%%%%%%%%%
%%%%%%%%%%%%%%%%%%%%%%%%%%%%%%%%%%%%%%%%%%%%%%%%%

\section{List of candidate states for $\nu=\frac{5}{2}$ ($\nu=\frac{1}{2}$) FQH state.\label{sec:candidatelist}}

\begin{table}
\caption{
Seven different candidates states for the $\nu=\frac{5}{2}$ FQH system with
number of edge branches, expected quasiparticle charge $e^*$ and
exponent $g$ for dominant tunneling quasiparticles, and exponent $g_e$
for electron operator.
Here, we have included the 2 right-moving
branches from the underlying $\nu=2$ IQH state.
The subscripts $L$ and $R$ indicate the left-moving and right-moving edge
branches.
Exponent $g$ generally seems to be
increasing with total number of edge branches. The listed electron operator
exponents $g_e$ ignore the $\nu=2$ IQH electrons. 
The two Majorana-gapped phases have a dominant $e^*=1/2$ quasiparticle in
addition to a quarter charge quasiparticle.
\label{tab:candidatelist}}
\begin{ruledtabular}
\begin{tabular}{ccccc}
state&\# of branches&$e^*$&\text{$g$}&$g_e$\\
\hline
$K=8$&$1_R+2_R$&1/4&0.125&$\infty$\\
\hline
Pfaffian&$\frac{3}{2}_R+2_R$&1/4&0.25&3\\
\hline
$331$&$2_R+2_R$&1/4&0.375&3\\ 
\hline
$U(1)\times SU(2)_2$&$\frac{5}{2}_R+2_R$&1/4&0.5&3\\ 
\hline
anti-Pfaffian&$1_R+\frac{3}{2}_L+2_R$&1/4&0.5&3\\
\hline
Majorana-gapped&$2_R+\frac{1}{2}_L+2_R$&1/4&0.5&3\\
{\shoveright {\ edge-rec. Pfaffian}}&&1/2&0.5 &\\
\hline
Majorana-gapped& $\frac{5}{2}_R+1_L$ & 1/4& 0.55-0.75&1.8-2.0\\
\shoveright{anti-Pfaffian}&& 1/2&0.5-0.7 &\\ 
\end{tabular}
\end{ruledtabular}
\end{table}

The Majorana-gapped phase at the edge of the anti-Pfaffian state is just one
of many possible edge states at filling fraction $\nu=\frac{5}{2}$.  Therefore, in
this section, we will review some known edge states for filling fraction
$\nu=\frac{5}{2}$. Or, to be more precise, for filling fraction
$\nu=\frac{1}{2}$ modulo completely filled Landau levels.

It is well known that, at a given filling fraction, FQH states may have many
different internal structures -- topological
orders.\cite{WNtop,Wtoprev,BWkmat1,R9002,WZclass,FZ9117,FK9169,FS9333} So it
is not clear which topological order describes a particular experimentally
observed $\nu=\frac{1}{2}$ ($\nu=\frac{5}{2}$) FQH state.  However, the following five topological orders
are simple and are more likely to describe the observed
$\nu=\frac{1}{2}$ ($\nu=\frac{5}{2}$) FQH states.
The five topological orders are:\\
(A) The electrons first pair into charge $2e$ bosons and the charge $2e$ bosons
then condense into the Laughlin state described by the following wave
function:
\begin{equation*}
\prod_{i<j}(Z_i-Z_j)^8
e^{-\frac{1}{4}\sum_i|Z_i|^2} .
\end{equation*}
The effective theory of this state has a form\cite{WNtop,BWkmat1,WZclass}
\begin{equation}
\cL=\sum_{IJ} K_{IJ}\frac{1}{4\pi}a_{I\mu}\prt_\nu a_{J\la}
\eps^{\mu\nu\la}
\label{eq:CS}
\end{equation}
with $K$ a $1\times 1$ matrix $K=8$.\\
(B) The charge $2e/3$ quasiparticles on top of the $\nu=1/3$ Laughlin state
condense into a second level hierarchical FQH state.\cite{YSS9219,BWkmat2,WZclass} The
effective theory of such a state is given by \Eq{eq:CS} with $K=\bpm 3&-2\\ -2&4\\
\epm$. Since
$ \bpm 3&1\\ 1&3\\ \epm= W \bpm 3&-2\\ -2&4\\ \epm W^T $
with $W=\bpm 1&1\\ 1&0\\ \epm$, such a state has the same topological order
as the $331$ double layer state.\cite{WZclass}
\\
(C) The FQH state proposed in {Ref.~} \onlinecite{MR9162} and described by the
Pfaffian wave function
\begin{equation}
\Psi_\text{pf}(\{z_i\})=\cA(\frac{1}{ z_1-z_2}\frac{1}{ z_3-z_4}\cdots)
\prod_{i<j}(z_i-z_j)^2
e^{-\frac{1}{4}\sum_i|z_i|^2}
\label{ising}
\end{equation}
where $\cA$ is the antisymmetrization operator.\\
(D) The anti-Pfaffian state, which is the particle-hole conjugate of the Pfaffian state.\cite{LHR0706,LRNF0707}\\
(E) The FQH state proposed in {Refs.~} \onlinecite{Wnab,BWnab} and described by the
wave function
\begin{equation}
%\Psi_{KM}
\Psi
(\{z_i\})=
[\chi_2(\{z_i\})]^2\prod_{i<j}(z_i-z_j)
e^{-\frac14\sum_i|z_i|^2}
\label{km}
\end{equation}
where $\chi_2(\{z_i\})$ is the fermion wave function of two filled Landau
levels. In the Appendix we provide a more detailed description of the edge theory of
this state.

Other topological orders at $\nu=\frac{1}{2}$ have more complicated internal
structures and are unlikely to appear.  For convenience, we will use $K=8$,
$331$, Pfaffian, anti-Pfaffian, and $U(1)\times SU_2(2)$ to
denote the above five topological orders respectively. 

The $K=8$ and the $331$ states are abelian FQH states, whose quasiparticles
all have abelian statistics.  The bulk low energy effective theories for the
two FQH state are given by $U(1)$ Chern-Simons (CS) theory,
Eq.~(\ref{eq:CS}). The edge excitations of 
the $K=8$ state are described by a single density mode (or more precisely, a
$U(1)_R$ Kac-Moody (KM) algebra, where the subscript $R$ indicates that the
excitations are right moving).  The number of low energy edge excitations for
the $K=8$ state is the same as one filled Landau level, as measured by the low
temperature specific heat.   Thus we say that the $K=8$ state has one branch
of edge excitations.  The edge excitations of the $331$ state are described by
two density modes (which form a $U(1)_R\times U(1)_R$ KM algebra).  Using the
similar definition in terms of specific heat, the $331$ state has two branches
of edge excitations.

The Pfaffian, anti-Pfaffian, and $U(1)\times SU_2(2)$ states are non-abelian states.  Some
of their quasiparticles have non-abelian statistics.  The edge excitations of
the Pfaffian state are described by a density mode (the $U(1)_R$ KM algebra) and a free
chiral Majorana fermion (the $\mathit{Ising}_R$ conformal field theory), or in other
word, by a $U(1)_R\times \mathit{Ising}_R$ conformal field theory (CFT).  Such an edge
state has one and a half branches of right-moving edge excitations as measured
by specific heat.  The edge excitations of the anti-Pfaffian state are described by
$U(1)_R\times U(1)_L\times \mathit{Ising}_L$ CFT.  The edge
excitations for the anti-Pfaffian state
have one and a half branches of left-moving edge excitations and one branch
of right-moving edge excitations.  For the  $U(1)\times SU_2(2)$ state, the
bulk effective theory is a $U(1)\times SU_2(2)$ CS theory and the edge
excitations are described by $U(1)_R\times SU_{2}(2)_R$ KM algebra.  The edge
state has two and a half branches of right-moving excitations.

The theory of edge excitations for both abelian and non-abelian FQH states
were well developed.\cite{Wtoprev,Wpcon,WWHopa,LHR0706,LRNF0707} In
Table~\ref{tab:candidatelist} we list the relevant results. Here $e^*$ is the
quasiparticles charge and $g$ is the exponent in the corresponding
quasiparticle Green's function: $\langle\psi_\text{qp}^\dag
\psi_\text{qp}\rangle \sim 1/t^g$. In terms of scaling dimensions
$\Delta$ the exponent $g$ is twice the scaling dimension of the
quasiparticle operator.

The results we find in this paper for the Majorana-gapped edge phases of the
anti-Pfaffian and edge-reconstructed Pfaffian states are
also included in the table.  Note that the anti-Pfaffian edge state and its Majorana-gapped edge state are two edge phases of the same anti-Pfaffian bulk
FQH state (and similarly for the Pfaffian edge states). For the
Majorana-gapped anti-Pfaffian phase we find that the exponent of the  quasiparticle Green's function
is non-universal; the exact value of $g$ depends on the interaction.
Nevertheless there are two dominant quasiparticles, one with $e^*=1/4$
and the exponent $g$ in the range $g\in[0.55-0.75]$, and one with $e^*=1/2$
and $g\in[0.5-0.7]$.

%%%%%%%%%%%%%%%%%%%%%%%%%%%%%%%%%%%%%%%%%%%%%%%%%%%%%%
%%%%%%%%%%%%%%%%%%%%%%%%%%%%%%%%%%%%%%%%%%%%%%%%%%%%%%

\section{Majorana-gapped phase of the anti-Pfaffian\label{sec:gappedantipf}}

This section hold the main results of our paper. We show in detail how
to calculate scaling dimensions of quasiparticle operators for the
anti-Pfaffian state in the presence of density-density
interactions. We identify a charge-transfer operator that can be
relevant. This operator is a product of a left-moving Majorana fermion
and a right-moving complex fermion. Condensation of this operator
gaps the left-moving Majorana mode and half of the right-moving
fermionic mode. 

In the new phase, dubbed `Majorana-gapped' phase, several
quasiparticle operators have become gapped as well, and we determine the
spectrum of ungapped quasiparticles. Next, we find the quasiparticle
with the lowest scaling dimension which is expected to dominate
tunneling. Finally, we consider the effect of impurities.

%%%%%%%%%%%%%%%%%%%%%%%%%%%%%%%%%%%%%%%%%%%%%%%%%%%

\subsection{Non-universality for non-chiral edges}

Fractional quantum Hall states which are described by fully chiral
edge theories are called `universal', because correlation function
exponents are independent of the the exact
microscopic details of e.g. the interaction between different edge
branches. 

This situation is no longer the case for FQH state described by an edge theory with branches moving in
opposite directions, i.e., a non-chiral edge. In this case the scaling
dimensions of operators depend on the exact form of the
interaction between the different edge branches, and a more detailed
analysis is required to predict the fate of e.g. tunneling
exponents. 

In some cases (i.e., for some regions in the space of all
possible interactions) the result is that the tunneling exponents are
indeed non-universal, in other cases the properties of the system are
dominated by a certain fixed point for which the tunneling exponents
do acquire universal values. In this sense one can construct a phase
diagram in `interaction-space'. 

Such an analysis typically focusses on two types of interactions:
density-density-interactions, which determine the scaling dimensions
of all quasiparticle operators, and charge-transfer
operators. Charge-transfer operators move charge between the different
edge branches, and as such it is also the mechanism through which
different edge branches equilibrate. Charge-transfer operators violate no
symmetry and are allowed to appear in the action. If a charge-transfer
operator has a relevant scaling dimension, the condensation of this
operator can lead to a different phase.

The effect of a charge-transfer operator can be more drastic than to merely
adjust values for tunneling exponents. For example in the $\nu=\frac{9}{5}$
state, it was shown\cite{KCW9963} that condensation of a charge-transfer
operator leads to a transition on the edge where a pair of counterpropagating
(previously gapless) edge modes becomes gapped; in the resulting
$\nu=\frac{9}{5}$ phase the single electron operator by itself is no longer
gapless.

%%%%%%%%%%%%%%%%%%%%%%%%%%%%%%%%%%%%%%%%%%%%%%%%%%%%%%%%%%

\subsection{$K$-matrix, action, electron operators, quasi-particles}

The edge theory of the anti-Pfaffian is described by one
(charge-neutral) Majorana
branch $\lambda$, and four (charge-carrying) bosonic branches $\phi_i$. In our setup three of the bosonic
branches are right-moving; the Majorana branch and the fourth bosonic
branch are left-moving. We adopt a basis where the left-moving branch
appears first, i.e., $(\phi_4,\phi_1,\phi_2,\phi_3)$; in this basis the $K$-matrix of
the bosonic modes is given by 
\begin{equation}
 K=\text{Diag}(-2,1,1,1).
\end{equation}
The action for this system is
\begin{multline}
S=\frac{1}{4\pi}\int dx d\tau \left[iK_{ij}\partial_x
\phi_i\partial_\tau
\phi_j+V_{ij}\partial_x\phi_i\partial_x\phi_j\right.\\
\left. +\lambda(v_\lambda\partial_x-\partial_\tau)\lambda\right].
\label{eq:antipfaction}
\end{multline}

The generic quasiparticle operator has a form of a bosonic vertex operator
$e^{i\vec l\cdot\vec \phi}$ times a Majorana operator. Majorana (CFT
primary field) operators are the identity operator
$\mathbbm{1}_\lambda$, the Majorana fermion operator $\lambda$ and the
spin operator $\sigma_\lambda$.  The charge and the bosonic contribution to the
(mutual) statistics of such a quasiparticle operator  can be determined from
the inverse of the $K$-matrix: \begin{equation}
\theta=\pi\, \vec l\cdot K^{-1}\cdot\vec l,\quad q=\vec t\cdot
K^{-1}\cdot \vec l,\quad \theta_{jk}=\pi\, \vec l_j\cdot
K^{-1}\cdot\vec l_k,
\label{eq:Kmatrixstat}
\end{equation}
where $\theta$ is the statistical phase, $q$ is the charge, $\vec
t=(1,1,\ldots,1)$ is the so-called charge vector and unit of charge is
$e=-|e|$. The Majorana branch is charge-neutral and commutes with the bosonic
branches. Its contribution to mutual statistics is\footnote{ For purposes of
finding the quasiparticle spectrum we do not need to consider mutual
statistics of two $\sigma$ operators.}
\begin{equation}
\frac{1}{\pi}\theta_{\lambda\lambda}=\pm
1,\qquad\frac{1}{\pi}\theta_{\lambda\sigma}=\pm\frac{1}{2},
\label{eq:Majoranastat}
\end{equation}
where we fix the sign to be $+1$ for right-moving branches and $-1$
for left-moving ones.

The quasiparticle spectrum is obtained by first identifying physical
electron operators, which have charge $e$ and fermionic
statistics. For the anti-Pfaffian state we are considering here, the physical
electron operators are $e_1=e^{i\phi_1}$,
$e_2=e^{i\phi_2}$, $e_3=e^{i\phi_3}$, $e_4=\lambda e^{-2i\phi_4}$, and
any combination of these $e_i$ with total charge $e$.

The remainder of the quasiparticle spectrum is formed by those quasiparticle
operators that are local with respect to all these electron operators, i.e.,
the phase induced by moving a quasiparticle around any electron operators
should be a multiple of $2\pi$. The allowed quasiparticles can
straightforwardly be found from these rules and are listed for convenience in
Table~\ref{tab:antipfspectrum}.

\begin{table}
\caption{Allowed quasiparticles in the $\nu=\frac{5}{2}$
anti-Pfaffian are labelled by four integers $m_j$ and a Majorana
sector ($\mathbbm{1}_\lambda$, $\lambda$, or
$\sigma_\lambda$). The corresponding vertex operators are $e^{i (m_1\phi_1+
m_2\phi_2+m_3\phi_3)}$ for the three right-moving branches and the
left-moving branch $\phi_4$ is explicitly listed in the table.
The quasiparticle charge
$q$ is also given.
\label{tab:antipfspectrum}}
\begin{ruledtabular}
\begin{tabular}{c|c|c}
$\lambda$-sector&$\phi_4$&$q$\\
\hline
$\mathbbm{1}_\lambda$&$e^{i m_4\phi_4}$&$m_1+m_2+m_3-\frac{m_4}{2}$\\
\hline
$\lambda$&$e^{im_4\phi_4}$&$m_1+m_2+m_3-\frac{m_4}{2}$\\
\hline
$\sigma_\lambda$&$e^{i( m_4-\frac{1}{2})\phi_4}$&$\frac{1}{4}+m_1+m_2+m_3-\frac{m_4}{2}$\\
\end{tabular}
\end{ruledtabular}
\end{table}

%%%%%%%%%%%%%%%%%%%%%%%%%%%%%%%%%%%%%%%%%%%%%%%%%%%%%%

\subsection{Calculating scaling dimension of quasiparticle
operators, boost parameters}

For the matrices $K$ and $V$ in the action, Eq.~(\ref{eq:antipfaction}), there
exist a (non-orthogonal) basis $\tilde \phi$ such that $K$ is a
pseudo-identity and $V$ is diagonal.  In such a basis, the scaling dimension
$\Delta$ of a quasiparticle operator  $e^{i\tilde l\cdot \tilde \phi}$ would
be given by $\Delta=\frac{1}{2}\tilde l^2$. In general, with $K$ given and
fixed, the scaling dimension of quasiparticle operators thus depends on the
precise form of the 4-by-4 matrix $V$ in the basis $\phi$.

A parametrization of the most generic density-density interaction $V$ requires
ten real parameters. With a suitable choice of parameters the scaling
dimension depends on only three of these parameters. This goes as follows. 

Let $M_1$ be the matrix that brings $K$ into pseudo-identity
form,
\begin{equation}
\tilde K=M_1^T K M_1=-\mathbbm{1}_{N^-\!}\oplus\mathbbm{1}_{N^+\!},
\end{equation}
 where $N^\pm$
is the number of positive/negative eigenvalues of $K$, in our case
$N^-\!=1$, $N^+\!=3$. Next we diagonalize $V$ with a matrix $M_2\in SO(N^-\!,N^+\!)$,
\begin{equation}
\tilde V=M_2^T M_1^T V M_1 M_2=\text{Diag}(v_4,v_1,v_2,v_3).
\end{equation}
Note that $M_2$ leaves $\tilde K$ invariant. Furthermore, $M_2$ can be
decomposed in a pure boost $B$ and a pure rotation $R\in SO(N^-\!)\times
SO(N^+\!)$, $M_2=B R$. 

In our case we use\footnote{
The extra rotation incorporated in our $M_1$ is added for later convenience.
}%
\begin{equation}
M_1=\left(
\begin{array}{llll}
 \sqrt{\frac{1}{2}} & 0 & 0 & 0 \\
 0 & 0 & \frac{1}{\sqrt{2}} & -\frac{1}{\sqrt{2}} \\
 0 & \sqrt{\frac{2}{3}} & \frac{1}{\sqrt{6}} & \frac{1}{\sqrt{6}} \\
 0 & \frac{1}{\sqrt{3}} & -\frac{1}{\sqrt{3}} & -\frac{1}{\sqrt{3}}
\end{array}
\right),
\label{eq:M1}
\end{equation}
and $B\in SO(1,3)$ is the familiar pure boost from the Lorentz-group,
\begin{equation}
 B=\left(
\begin{array}{llll}
 \gamma  & \beta_1 \gamma  & \beta_2 \gamma  & \beta_3 \gamma  \\
 \beta_1 \gamma  & \frac{\beta_1^2 \gamma ^2}{\gamma +1}+1 & \frac{\beta_1 \beta_2 \gamma
   ^2}{\gamma +1} & \frac{\beta_1 \beta_3 \gamma ^2}{\gamma +1} \\
 \beta_2 \gamma  & \frac{\beta_1 \beta_2 \gamma ^2}{\gamma +1} & \frac{\beta_2^2 \gamma ^2}{\gamma
   +1}+1 & \frac{\beta_2 \beta_3 \gamma ^2}{\gamma +1} \\
 \beta_3 \gamma  & \frac{\beta_1 \beta_3 \gamma ^2}{\gamma +1} & \frac{\beta_2 \beta_3
   \gamma ^2}{\gamma +1} & \frac{\beta_3^2 \gamma ^2}{\gamma +1}+1
\end{array}
\right),
\label{eq:boost}
\end{equation}
with $\gamma=1/\sqrt{1-(\beta_1^2+\beta_2^2+\beta_3^2)}$. An explicit
specification of $R$ is not required at this point.

What is important from this decomposition is that the scaling dimension of operator $e^{i\vec
l\cdot \vec \phi}=e^{i\tilde l\cdot\tilde\phi}$ is now given by $\Delta=\frac{1}{2}\vec
l\cdot \tilde \Delta\cdot\vec l$, where
\begin{equation}
\tilde \Delta=M_1 B^2 M_1^T.
\end{equation}
In our case, $V$ is parametrized by four (eigenvalues $v_j$) plus
three (rotation $R\in SO(3)$) plus three (boost parameters $\beta_j$)
equalling a total of ten parameters. But the scaling dimension depends only
on the three boost parameters $\beta_j$: one parameter for each pair
of counterpropagating edge modes. 
The scaling dimension in the bosonic sector of any
quasiparticle operator thus becomes a function of a vector $\vec
\beta=(\beta_1,\beta_2,\beta_3)$ inside the unit 3D-ball $\beta^2<1$
(cf. $|v|<c$).

There is also a contribution to the scaling dimension from the
Majorana sector, 
\begin{equation}
\Delta_\lambda=\frac{1}{2},\qquad\Delta_{\sigma}=\frac{1}{16},
\end{equation}
 which simply needs to be added to the bosonic scaling dimension of a
quasiparticle operator to obtain the total scaling dimension.  Here we note
that the density-density interactions  between the Majorana sector and the
boson sectors are always irrelevant and we ignore those interactions in our
calculations of scaling dimensions.  We will consider other interactions
between the Majorana sector and the boson sectors below.

%%%%%%%%%%%%%%%%%%%%%%%%%%%%%%%%%%%%%%%%%%%%%%%%%%%%%%%%%%%%%%%%%%

\subsection{Majorana mode becomes gapped through `null' charge-transfer operator}

\begin{figure*}
\includegraphics[width=12cm]{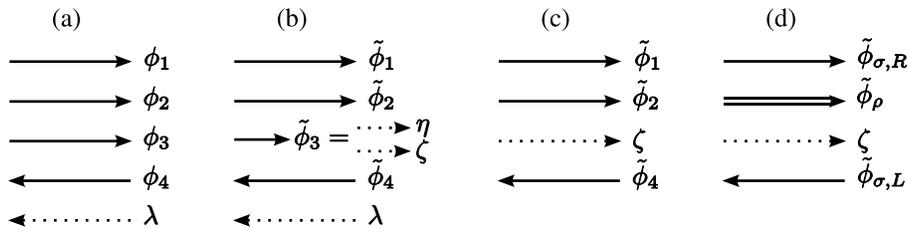}
\caption{Different representations of the gapless branches in the
anti-Pfaffian and Majorana-gapped phases. (a) In the basis $\phi$ the anti-Pfaffian has three
right-moving bosonic branches (solid lines) and two left-moving modes,
one bosonic and one Majorana mode (dotted line). (b) After a
basistransformation to basis $\tilde \phi$ the right-moving charge-neutral mode $\tilde
\phi_3$ can be expressed as two right-moving Majorana modes $\eta$ and
$\zeta$. (c) In the Majorana-gapped phase, the modes $\lambda$ and
$\eta$ acquire a gap and are dropped. (d) An additional
basistransformation explicitly separates the charge mode, $\tilde \phi_\rho$
(double solid line), from
the neutral modes ($\tilde \phi_{\sigma,L/R}$).
 \label{fig:branches}}  
\end{figure*}

Now that we know how to calculate scaling dimensions, we can probe
$\vec \beta$-space for charge-transfer operators with low scaling
dimension. Charge-transfer operators are total charge zero operators,
which typically move electrons between the different branches.
Since they violate no symmetry or conservation, charge-transfer
operators are in principle allowed to appear in the action,
Eq.~(\ref{eq:antipfaction}), and if relevant (in RG sense) can cause a transition to another phase.

In
$\vec\beta$-parameter-space surfaces of constant scaling dimension
typically are of ellipsoidal shape, for example
\begin{equation}
 \Delta_{+e_2- e_3}=1+\frac{2 \beta_1^2}{1-\beta^2},\quad \Delta_{+e_1-e_4}=\frac{2
\left(\beta_2+\frac{\beta_3}{\sqrt{2}}+\sqrt{3}\right)^2}{3
\left(1-\beta ^2\right)},
\label{eq:Deltaexamples}
\end{equation}
where $+e_i- e_j$ stands for the combination of an $e_i$
creation and an $e_j$ destruction operator, which transfers charge $e$
between branches $i$ and $j$. 

Regions inside parameter space where charge-transfer operators are relevant ($\Delta<2$) are
thus ellipsoids inside the unit ball $\beta^2<1$.

One charge transfer operator we are particularly interested in is
$\lambda e^{i(2\phi_4-\phi_1+\phi_2+\phi_3)}\equiv\hat n$, that is, the operator
which simultaneously destroys $e_1$ and $e_4$ and creates $e_2$ and
$e_3$. Its scaling dimension is
\begin{equation}
\Delta_{-e_1-e_4+e_2+e_3}=1+\frac{3\left(\sqrt{\frac{2}{3}}-\beta_3\right)^2}{1-\beta^2}.
\end{equation}
There is a whole \emph{disc} in $\vec \beta$-space for which the
scaling dimension of $\hat n$ is
identically $1$, namely when  $\beta_3=\sqrt{2/3}$ (with
$\beta_1^2+\beta_2^2<1/3$). 

On this disc the bosonic part of the
operator $\hat n$ has the scaling dimensions and statistics of a charge-neutral
right-moving complex fermion, which we write as a combination of two
Majorana fermions $\eta$ and $\zeta$, such that $\hat
n=\lambda(\eta+i\zeta)$. Note that $\hat n$ resembles the
`neutral null vector' from the $\nu=\frac{9}{5}$ FQH case, as in it
is a zero-charge operator with equal left and right conformal dimensions, $h=\bar h$.

We are now approaching the step where the Majorana mode acquires a
gap. Clearly this is a key ingredient in our procedure. But the
argument of gapping itself is almost trivial: consider a system with two
counterpropagating fermions $\psi_1$ and $\psi_2$, with dispersion
relation $E_{1/2}(k)=\pm v k$, which has gapless excitations at zero
energy; adding a coupling $\sum_x
\Gamma \psi^\dag_1(x)\psi_2(x)+\text{H.c.}$ changes the dispersion to
$E_\pm(k)=\sqrt{{\Gamma}^2+(vk)^2}$ and opens up a gap at zero energy. 

Let us assume the interaction $V$ is such that the operator $\hat n$ is
relevant, and include $\hat n$ and its Hermitian conjugate in the
action Eq.~(\ref{eq:antipfaction}) with a \emph{constant} coupling $\Gamma$, i.e., we 
are not considering disorder at this point. Then the effect of this
term,..
$\Gamma(\hat n+\hat n^\dag)=2\Gamma\lambda\eta$,
 is that the
counterpropagating Majorana modes 
$\lambda$ and $\eta$ become gapped whereas $\zeta$ is left
untouched. \emph{In other words the left-moving
Majorana mode and a right-moving bosonic mode disappear and a
right-moving Majorana mode emerges.}

\begin{figure}
\includegraphics[width=6cm]{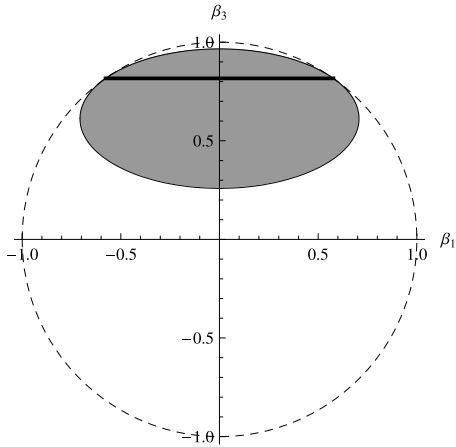}
\caption{Shown is the cross-section, at $\beta_2=0$, of the parameter space unit ball
$|\vec \beta|^2<1$. The scaling dimension of the `null'
operator $\hat n$, the charge-transfer operator which induces the
transition to the Majorana-gapped phase, is identically one at the
disc $\beta_3=\sqrt{2/3}\approx0.82$ (indicated by thick black
line). This null operator $\hat n$ is relevant in a substantial volume
of $\vec \beta$ parameter space; the grey area indicates the
ellipsoidal volume where $\hat n$ had scaling dimension
$\Delta\le3/2$, i.e., the region where $\hat n$ is relevant even in
the presence of disorder.
\label{fig:ellipse32}
}
\end{figure}

 Figure~\ref{fig:ellipse32} shows the volume of
parameter space in which $\hat n$ is a relevant operator. A schematic
representation of the branches and different bases before and after
Majorana-gapping is given in Fig.~\ref{fig:branches}.

Before we continue to determine the quasiparticle spectrum in the
Majorana-gapped system we would like to note that the operator $\hat
n$ is not unique; due to the permutation symmetry between branches
$\phi_1$, $\phi_2$ and $\phi_3$ there is a total of three such
operators $\hat n$.

%%%%%%%%%%%%%%%%%%%%%%%%%%%%%%%%%%%%%%%%%%%%%%%%%%%%%%%%%%%%%%%%%%%%%%%

\subsection{Quasiparticle spectrum in gapped
system}

With the gapping of the $\lambda$ Majorana fermion, some of the
quasiparticle operators in the original spectrum, Table~\ref{tab:antipfspectrum}, have
likely developed a gap as well and have disappeared from the
low-energy spectrum. 
For the Majorana-gapped phase we would like to find out (i) which
quasiparticles have survived the gapping and (ii) which of these
survivors have the
lowest scaling dimension and are thus expected to dominate e.g. tunneling
processes. 

To obtain the quasiparticle spectrum we follow the same
procedure as for any other $\chi$LL FQH edge system: we identify the physical electron
operators and determine those quasiparticles which are single-valued
with respect to the electron operators. The non-standard part is how
to remove the degree of freedom associated with the now-gapped
$\lambda$ Majorana fermion and insert the now-emerged $\zeta$ Majorana fermion.

Setting $\beta_3=\sqrt{2/3}$, the following steps will find the
quasiparticle spectrum for arbitrary $\beta_1$ and $\beta_2$
(${\beta_1^2+\beta_2^2<1/3}$). To illustrate the procedure we will use
the example where $\beta_1=0=\beta_2$, for which the intermediate
basis-dependent values are relatively simple.

We transform the basis from $\phi$ to $\tilde \phi$ such that $\tilde K$ is
the pseudo-identity, $\tilde V$ is diagonal, and $\hat n=\lambda e^{+i\tilde\phi_3}$, which is achieved by $M_1$ and
$B$, Eqs.~(\ref{eq:M1}) and (\ref{eq:boost}), and $R=\text{Diag}(1,1,-1,-1)$.
As far as the electron operators go, $e_4$ contains a $\lambda$
operator and hence becomes gapped, so we drop $e_4$ from the spectrum.
The three remaining physical electron operators have the following
form in the basis $\tilde \phi$,
\begin{align}
e_1&=e^{i\text{$\bigl($}+\frac{\sqrt{2}}{\sqrt{3}}\tilde\phi_4+\frac{\sqrt{2}}{\sqrt{3}}\tilde\phi_2-\tilde\phi_3\text{$\bigr)$}},\\ 
e_2&=e^{i\text{$\bigl($}-\frac{\sqrt{2}}{\sqrt{3}}\tilde\phi_4+\frac{1}{\sqrt{2}}\tilde\phi_1-\frac{1}{\sqrt{6}}\tilde\phi_2+\tilde\phi_3\text{$\bigr)$}},\\
e_3&=e^{i\text{$\bigl($}-\frac{\sqrt{2}}{\sqrt{3}}\tilde\phi_4-\frac{1}{\sqrt{2}}\tilde\phi_1-\frac{1}{\sqrt{6}}\tilde\phi_2+\tilde\phi_3\text{$\bigr)$}}.	
\end{align}

For the operator $e^{+i\tilde\phi_3}=\eta+i\zeta$ we expect that
gapping will get rid of the $\eta$-part and effectively leave a
Majorana operator $\zeta$ times some overall phase:
$e^{i\tilde\phi_3}\to\zeta$. Note that this includes all three
electron operators for which we thus make the identification
\begin{align}
e_1&\simeq
\zeta\,e^{i\text{$\bigl($}+\frac{\sqrt{2}}{\sqrt{3}}\tilde\phi_4+\frac{\sqrt{2}}{\sqrt{3}}\tilde\phi_2\text{$\bigr)$}},\\
e_2&\simeq
\zeta\,e^{i\text{$\bigl($}-\frac{\sqrt{2}}{\sqrt{3}}\tilde\phi_4+\frac{1}{\sqrt{2}}\tilde\phi_1-\frac{1}{\sqrt{6}}\tilde\phi_2\text{$\bigr)$}},\\
e_3&\simeq
\zeta\,e^{i\text{$\bigl($}-\frac{\sqrt{2}}{\sqrt{3}}\tilde\phi_4-\frac{1}{\sqrt{2}}\tilde\phi_1-\frac{1}{\sqrt{6}}\tilde\phi_2\text{$\bigr)$}}.
\end{align}

Next, we look for quasiparticles which are single-valued with respect
to these three electron operators, with the generic form $e^{i(\tilde
l_4\tilde\phi_4+\tilde l_1\tilde\phi_1+\tilde l_2\tilde \phi_2)}$ times
a $\zeta$-sector Majorana operator $\mathbbm{1}_\zeta$, $\zeta$, or 
$\sigma_\zeta$.

Solving for allowed $\tilde l_4$, $\tilde l_1$ and $\tilde l_2$ now is
a computationally trivial task of solving a set of three linear
equations given by the mutual statistics equations,
Eqs.~(\ref{eq:Kmatrixstat})~and~(\ref{eq:Majoranastat}).
The resulting expressions for the $\tilde l_j$ will involve various
square roots,
 which tend to become more ugly for
generic values for $\beta_1$ and $\beta_2$. However, the mutual statistics
equations are basis-invariant, and hence can be solved in any
basis. As it turns out, these equations become really simple in the
original $\phi$ basis.

Making the identifications $\zeta\simeq e^{+i\tilde\phi_3}$ and
$\sigma_\zeta\simeq e^{+i\frac{1}{2}\tilde\phi_3}$ we can transform
back to the original basis $\phi$. In this basis, the generic
quasiparticle operator has the form $e^{i(l_4\phi_4+l_1\phi_1+l_2\phi_2+l_3\phi_3)}$. Single-valuedness with the three electron
operators forces $l_1$, $l_2$, and $l_3$ to be integers $m_j$; $l_4$ is determined by
the constraint
\begin{equation}
-l_4-l_1+l_2+l_3=\begin{cases}
0&\text{for }\mathbbm{1}_\zeta\text{-sector,}\\
1&\text{for }\zeta\text{-sector,}\\
\frac{1}{2}&\text{for }\sigma_\zeta\text{-sector,}
\end{cases}
\end{equation}
which will assure the appropriate coefficient of $\tilde\phi_3$ for
each Majorana sector in the $\tilde\phi$ basis.

So the result is that the quasiparticle spectrum in the
Majorana-gapped phase can be labelled by three integers $m_1$, $m_2$
and $m_3$, and a Majorana $\zeta$-sector, as shown in
Table~\ref{tab:gappedspectrum}. Note that these expressions are
\emph{independent} of $\beta_1$ and 
$\beta_2$. The charge is given by $q=l_1+l_2+l_3-\frac{1}{2}l_4$. 

Even the scaling dimension can
be calculated in the $\phi$ basis for all $\beta_1$ and $\beta_2$, however,
in the $\sigma_\zeta$ sector the calculation $\frac{1}{2}\tilde l_3^2$
would assign a scaling dimension of $\frac{1}{8}$ to the
$\sigma$-operator; we know this should be $\frac{1}{16}$ for a
Majorana $\sigma$ operator, and so scaling
dimension calculations need to be corrected for
this. This stems from the identification $\sigma_\zeta\simeq
e^{+i\frac{1}{2}\tilde\phi_3}$, which gives the correct statistics and
is valid for the purpose of enumerating the quasiparticle spectrum, but may not be true as operator equality.

\begin{table}
\caption{Quasiparticle spectrum in the Majorana-gapped
phase. Quasiparticles are identified by three integers $m_1$, $m_2$,
$m_3$ and a $\zeta$-Majorana sector. The corresponding vertex operator
are easiest expressed in the \emph{ungapped} basis $\phi$, where the
$\phi_1$, $\phi_2$ and $\phi_3$ contributions are still $e^{i
(m_1\phi_1+ m_2\phi_2+m_3\phi_3)}$. The three $m_j$ and the Majorana
sector fix the coefficient of $\phi_4$, as shown in the table.
The quasiparticle charge $q$ is listed, as well as a correction to the
scaling dimension, $\Delta_\text{cor}$, as
explained in the text.
\label{tab:gappedspectrum}}
\begin{ruledtabular}
\begin{tabular}{c|c|c|c}
$\zeta$-sector&$\phi_4$&$q$&$\Delta_\text{cor}$\\
\hline
$\mathbbm{1}_\zeta$&$e^{i (-m_1+m_2+m_3)\phi_4}$&$\frac{3}{2}m_1+\frac{1}{2}m_2+\frac{1}{2}m_3$&$0$\\
\hline
$\zeta$&$e^{i
(-m_1+m_2+m_3-1)\phi_4}$&$\frac{1}{2}+\frac{3}{2}m_1+\frac{1}{2}m_2+\frac{1}{2}m_3$&$0$\\ 
\hline
$\sigma_\zeta$&$e^{i( -m_1+m_2+m_3-\frac{1}{2})\phi_4}$&$\frac{1}{4}+\frac{3}{2}m_1+\frac{1}{2}m_2+\frac{1}{2}m_3$&$-\frac{1}{16}$\\
\end{tabular}
\end{ruledtabular}
\end{table}

%%%%%%%%%%%%%%%%%%%%%%%%%%%%%%%%%%%%%%%%%%%%%%%%%%%%%%%%%%%%%%%%

\subsection{Dominant quasiparticles in gapped system, charge
separation}

Having determined the quasiparticle spectrum, we now look for the most
dominant quasiparticles.

As far as (non-)universality goes, the gapping of the pair of Majorana
modes has removed one pair of counterpropagating bosonic modes from
the system, and with it one boost-parameter. Two
counterpropagating pairs remain with corresponding boost parameters
$\beta_1$ and $\beta_2$.

And so in principle we now have to repeat our procedure of looking for
dominant charge-transfer operators on the disc
$\beta_1^2+\beta_2^2<1/3$. However, so far we have considered the
most general density-density interaction $V$. We expect the
interaction to show traces of the underlying Coulomb interaction;
especially, we expect that there will be a single charge mode which will separate itself from
the other (neutral) modes.

Here we will consider the limit where the charged mode is completely
separated from the neutral modes. This decouples one of the
right-moving bosonic modes from the left-moving one and eliminates one
boost parameter. The condition for charge-separation is
$\beta_2=(\sqrt{2}\beta_3-\sqrt{3})/4=-1/(4\sqrt{3})$.
The one remaining boost parameter is $\beta_1$, with
$|\beta_1|<\sqrt{5}/4=\sqrt{1-\beta_2^2-\beta_3^2}$.

\begin{figure}
\includegraphics[width=8.6cm]{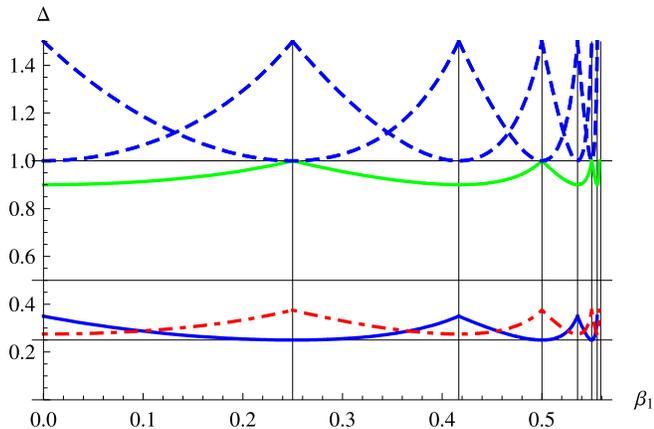}
\caption{
The scaling dimension of several important quasiparticles in the Majorana-gapped
phase as function of boost parameter $\beta_1$. Plotted are
charge-transfer operators with scaling dimension $\Delta<3/2$ (dashed
blue lines), electron operator with lowest scaling dimension (solid
green line), and two operators with lowest scaling dimension of all
operators: a charge $q=1/4$ quasiparticle (dashed-dotted red
line) and a charge $q=1/2$ quasiparticle (solid blue line). Notice
that there is a pattern which repeats itself for $\beta_1>0.25$,
turning into a series with shrinking width as
$\beta_1$ approaches its maximum allowed value
$\beta_1=\sqrt{5}/4\approx0.56$. Note that there are many
quasiparticle operators with scaling dimensions smaller than $1.5$
that we did not include on this graph.
\label{fig:Deltaline}}
\end{figure}

So we continue our analysis of scaling dimensions of operators on the
line $\beta_1$. A plot of scaling dimensions for several quasiparticle
operators is given in Fig~\ref{fig:Deltaline}. 
Upon closer inspection though, there is some regularity in the
spectrum. For instance, charge-transfer operators can have a minimal scaling
dimension of one, which is obtained for
$\beta_1=0,\pm\frac{1}{4},\pm\frac{5}{12},\pm\frac{1}{2},\pm\frac{15}{28},\ldots$
which appears to form an on-going series, and in between such points
the same `spectrum' of scaling dimensions is repeated. 

So it seems we only need to consider the interval
$0\le\beta_1\le1/4$. At $\beta_1=0$ the most dominant
quasiparticle operator is a  charge $q=1/4$ $\sigma_\zeta$-sector operator,
with scaling dimension $\Delta=0.275$. Upon increasing $\beta_1$
the scaling dimension increases monotonically to a value of
$\Delta=0.375$ at $\beta_1=1/4$. At $\beta_1=1/4$ the
quasiparticle operator with the lowest scaling dimension is a
charge $q=1/2$ $\mathbbm{1}_\zeta$-sector operator with $\Delta=0.25$. Its scaling dimension
increases monotonically in the opposite direction, reaching a maximum
at $\beta_1=0$ of $\Delta=0.35$.

It is tempting to suggest that the charge-transfer operator with the
smallest scaling dimension will dominate and fix the system to be either
at the $\beta_1=0$ or at the $\beta_1=1/4$ point. However,
since both charge-transfer operators have scaling dimension between 1
and 3/2 on the interval, they are both relevant, and it depends on the
strength of the coefficient if one dominates over the
other. Similarly, in our analysis we cannot single out a most dominant quasiparticle,
it is simply too close to call. In that sense we find the
Majorana-gapped charge-separated phase to be non-universal: there are two
dominant quasiparticles, with charges of $1/4$ and $1/2$ and scaling
dimensions ranging between $0.25$ and $0.375$.

%%%%%%%%%%%%%%%%%%%%%%%%%%%%%%%%%%%%%%%%%%%%%%%%%%%%%%%%%%%%%%%%%%%%%%

\subsection{Only strong interaction leads to Majorana-gapped phase}

Having identified the Majorana-gapped phase, we can ask what kind of
interaction will lead to such a phase. In the Majorana-gapped
charge-separated phase the interaction is characterized by 5
remaining parameters: if we pick $\tilde \phi_2$ as the charged mode these are
the three $\tilde V$ eigenvalues $v_4$, $v_1$ and $v_3$, an angle
$\alpha$ for rotations between branches $\tilde \phi_1$ and $\tilde
\phi_3$, and the boost parameters $\beta_1$.

A crucial ingredient for the
Majorana-gapped phase is to include the two filled (lowest) Landau
level modes. If these two modes are spatially well-separated on the
edge from the inner two modes one would expect the interactions to be
small between these blocks of edge branches. We find that the
interaction required for the Majorana-gapped phase is such that this
kind of separation of two modes is not possible: all right-moving branches
have to interact with the left-moving branch with similar strength.

%%%%%%%%%%%%%%%%%%%%%%%%%%%%%%%%%%%%%%%%%%%%%%%%%%%%%%%%%%%%%%%%%%%%

\subsection{Disorder: localization instead of gapping}

The assumption we made so far was that the charge-transfer coupling
strength $\Gamma$ was uniform along the edge. A more realistic assumption would
be to consider $\Gamma=\Gamma(x)$ to be fluctuating with position due to random disorder. 
Also, with disorder we do not need to worry about momentum mismatch
between different edge modes.

With disorder present, we expect instead of the gapping of the pair of left
and right moving Majorana modes, $\lambda$ and $\eta$, that they will become
localized.  Here we will assume that the localized modes do not contribute to
the tunneling between edges. In particular, they do not affect the value of
exponent $g$.  So as long as the calculation of $g$ is concerned, we treat the
localized modes as if the are gapped.  Thus the above calculation of $g$ also
applies to the disordered edge with localization.

%%%%%%%%%%%%%%%%%%%%%%%%%%%%%%%%%%%%%%%%%%%%%%%%%%%%%%%%%%%%%%%%%%
%%%%%%%%%%%%%%%%%%%%%%%%%%%%%%%%%%%%%%%%%%%%%%%%%%%%%%%%%%%%%%%%%%

\section{Majorana-gapped phase of the edge-reconstructed Pfaffian
state\label{sec:gappedpf}}

We now apply the same mechanism of Majorana gapping on a different
state: the edge-reconstructed Pfaffian state. By itself the Pfaffian \cite{MR9162}
state is fully chiral and gapping of pairs of counterpropagating modes
cannot occur. However, the edge might be unstable towards 
edge reconstruction \cite{CW9427}. Edge reconstruction effectively
adds pairs of counterpropagating charged bosonic modes to the edge. Here we
will analyze the state in which edge reconstruction has introduced one
such pair of edge modes to the Pfaffian state.

In the edge-reconstructed Pfaffian there are three bosonic modes
$\phi_1$, $\phi_2$ and $\phi_3$, and one neutral Majorana mode
$\lambda$. The left-moving branch is $\phi_1$, the other branches are
right-moving. The $K$-matrix is $K=\text{Diag}(-1,1,2)$. Electron
operators are $e_1=e^{i\phi_1}$, $e_2=e^{i\phi_2}$ and $e_3=\lambda e^{-2i\phi_3}$.

The `null' operator $\lambda
e^{i(2\phi_1+\phi_2+2\phi_3)}$ is a charge-transfer operator with
equal left and right conformal dimensions
$h=\bar h=\frac{1}{2}$. Introducing boost parameters $\vec \beta=
(\beta_1,\beta_2)$, similar to Eq.~(\ref{eq:boost}), we can parametrize scaling
dimensions of quasiparticle operators. The scaling dimension of $\lambda
e^{i(2\phi_1+\phi_2+2\phi_3)}$ becomes one at the point
$\beta_1=-1/2$, $\beta_2=-1/\sqrt{2}$.

We perform a basistransformation from $\phi$ to $\tilde\phi$; in this
basis $\tilde K=\text{Diag}(-1,1,1)$, $\lambda
e^{i(2\phi_1+\phi_2+2\phi_3)}=\lambda e^{i\tilde \phi_1}$, and
the $\tilde\phi_2$ branch carries all the charge. If the null
operator is relevant it will gap the right-moving Majorana mode
$\lambda$ and half of the left-moving bosonic mode $\tilde\phi_1$
leaving a left-moving Majorana mode $\zeta$.

In the Majorana-gapped phase, the gapless physical electron operators are $e_1=\zeta
e^{i(\sqrt{2}\tilde\phi_2)}$ and
$e_2=e^{i(\sqrt{2}\tilde\phi_2+\tilde\phi_3)}$; $e_3$ acquires a
gap. The quasiparticle
spectrum can be labelled by two integers $m_1$ and $m_2$ and a $\zeta$
Majorana sector, as follows, with charge and scaling dimension included:
\begin{multline}
\textstyle\mathbbm{1}_\zeta\text{-sector}:\quad
e^{i\frac{m_1}{\sqrt{2}}\tilde\phi_2+(m_2-m_1)\tilde\phi_3}
,\quad q=\frac{m_1}{2},\\
\textstyle
\Delta=\frac{m_1^2}{4}+\frac{(m_2-m_1)^2}{2},
\end{multline}
\begin{multline}
\textstyle\zeta\text{-sector}:\quad\zeta\,
e^{i\frac{m_1}{\sqrt{2}}\tilde\phi_2+(m_2-m_1)\tilde\phi_3}
,\quad q=\frac{m_1}{2},\\
\textstyle
\Delta=\frac{m_1^2}{4}+\frac{(m_2-m_1)^2}{2}+\frac{1}{2},
\end{multline}
\begin{multline}
\textstyle\sigma_\zeta\text{-sector}:\quad
\sigma_\zeta\,e^{i\frac{m_1+\frac{1}{2}}{\sqrt{2}}\tilde\phi_2+(m_2-m_1-\frac{1}{2})\tilde\phi_3}
,\quad q=\frac{m_1+\frac{1}{2}}{2},\\
\textstyle
\Delta=\frac{(m_1+\frac{1}{2})^2}{4}+\frac{(m_2-m_1-\frac{1}{2})^2}{2}.
\end{multline}

Note that here the Majorana-gapping effectively removes all pairs of
counterpropagating bosonic modes that existed before. Hence the
scaling dimensions of all operators becomes fixed. In other words,
there is no remaining boost parameter degree of freedom.

In the Majorana-gapped edge-reconstructed there are \emph{three}
operators with smallest scaling dimension $\Delta=\frac{1}{4}$: one
charge $q=1/2$ operator ($m_1=m_2=1$ in $\mathbbm{1}_\zeta$-sector), and
two charge $q=1/4$ operators ($m_1=0$, $m_2=0,1$ in the $\sigma_\zeta$-sector).
The electron operator with smallest scaling dimensions has $\Delta=\frac{3}{2}$.

%%%%%%%%%%%%%%%%%%%%%%%%%%%%%%%%%%%%%%%%%%%%%%%%%%%%%%%%%%%%%%%%%
%%%%%%%%%%%%%%%%%%%%%%%%%%%%%%%%%%%%%%%%%%%%%%%%%%%%%%%%%%%%%%%%%

\section{Discussion\label{sec:discussion}}

\subsection{Tunneling through bulk in new edge phase}

To detect the phase transition to the Majorana-gapped phase
on the edge of the anti-Pfaffian state one would have to observe a
change in quasiparticle 
tunneling exponent $g$. This presents a dilemma: even though $g$
itself is an intrinsic property of the edge, a  measurement of $g$
requires the quasiparticle to 
tunnel through the bulk. But in the bulk the phase transition does not
occur, so is it even possible for the quasiparticle to tunnel? 
We assume that edge quasiparticles in the Majorana-gapped phase
can indeed tunnel through the bulk and we do not run into obvious
inconsistencies (of e.g. having a quasiparticle charge on the edge
which does not exist in the bulk). Whether or not this assumption is
fully justified is not yet understood.

\subsection{Charge transfer in the bulk}

We like to stress that that operator  $\lambda
e^{i(2\phi_4-\phi_1+\phi_2+\phi_3)}$ not only appears in the edge effective
Hamiltonian, the corresponding operator also appears in the 2D bulk effective
Hamiltonian for the 2D anti-Pfaffian state.  Such a bulk operator transfers
charges between different condensates (note that the anti-Pfaffian state is
formed by several condensates: the spin-down electrons in the first Landau
level, the spin-up electrons in the first Landau level, the spin-up electrons
in the second Landau level, etc).  With such an operator present in the bulk
Hamiltonian, one naturally expects that the 2D anti-Pfaffian state has
$1_L+(5/2)_R$ branches of the edge excitations.  The $(3/2)_L+3_R$  branches
of the edge excitations proposed in {Refs.~} \onlinecite{LHR0706,LRNF0707} can
be viewed as a result of edge reconstruction of the $1_L+(5/2)_R$
edge.\cite{CW9427}

\subsection{Effects of spin conservation}

So far we have ignored the effect of spin conservation.  In the presence of
magnetic field, the $z$-component of spin $S_z$ is still conserved.  By
examine the spin quantum number of the charge transfer operator $\lambda
e^{i(2\phi_4-\phi_1+\phi_2+\phi_3)}$ in the anti-Pfaffian state, we find that
it carries $S_z=1$.  Therefore, the $S_z$ conservation prevents $\lambda
e^{i(2\phi_4-\phi_1+\phi_2+\phi_3)}$ from appearing in the edge Hamiltonian.
In this case the Majorana-gapped phase for the  anti-Pfaffian state cannot
appear.  Thus to have the Majorana-gapped phase for the  anti-Pfaffian state
we either need to break the $S_z$ conservation, or to consider the $\nu=9/2$
anti-Pfaffian state where there exists a  charge transfer operator which carries
$S_z=0$.

The charge transfer operator for the edge reconstructed Pfaffian state has
$S_z=0$. Thus the $S_z$ conservation will not prevent the appearance of the
Majorana-gapped phase.  The Majorana-gapped phase is more likely to appear for
edge reconstructed Pfaffian state.

\subsection{Determining the true nature of the $\nu=\frac{5}{2}$ state}

Glancing at Table~\ref{tab:candidatelist} we have to conclude that the
quest to determine the nature of the observed $\nu=\frac{5}{2}$ FQH
state is far from over. The first experimental results
\cite{RMMKPW0803.3530,DHUSM0802.0930} suggest that likely $e^*=1/4$ and $g=0.5$.
If these are confirmed to be the correct values we can scratch a few
candidates off the list; but we would not be able to distinguish between the
anti-Pfaffian, $U(1)\times SU_2(2)$ and Majorana-gapped
edge-reconstructed Pfaffian states. Electron tunneling is expected to
be the same for these three states as well.

 Additional measurements which would probe the
number of left- and right-moving edge branches would be required to
settle this issue. For instance a thermal Hall conductance \cite{KF9732} measurement
distinguishes between the three states with $e^*=1/4$ and $g=0.5$.

As far as the presence of non-abelian statistics goes the prospect is
somewhat brighter, as five out of the seven candidate states are non-abelian.
Furthermore the non-abelian statistics is carried by similar Ising spin
fields in all these cases, hence experimental setups based on interference should give
qualitatively similar results.

\subsection{Summary}

In the paper, we studied the effect of a charge transfer process described by
neutral bosonic operators in the $\nu=\frac{5}{2}$ anti-Pfaffian state and
edge reconstructed Pfaffian state.  On the edge, such operators have a form
$\lambda e^{i(2\phi_4-\phi_1+\phi_2+\phi_3)}$ or $\lambda
e^{i(2\phi_1+\phi_2+2\phi_3)}$.  Such operators transfer charges between edge
branches and create/annihilate a Majorana fermion $\lambda$.  The operator
respects all the symmetries and is local with respect to all the electron
operators.  Thus such an operator is allowed in the effective edge
Hamiltonian.  We find that, for a certain range of interactions between the
edge branches, the operators $\lambda e^{i(2\phi_4-\phi_1+\phi_2+\phi_3)}$ or
$\lambda e^{i(2\phi_1+\phi_2+2\phi_3)}$ represent relevant perturbations.  The
effect of such a relevant perturbation opens up a gap for a pair of left and
right moving Majorana fermion modes.  

For the anti-Pfaffian state, before the 1D gapping transition at the edge, the
state has 3/2 branches of left-movers and 3 branches of right-movers.  After
the gapping transition, the same 2D anti-Pfaffian state has 1 branch of
left-movers and 2 and 1/2 branches of right-movers.  For the edge
reconstructed Pfaffian state, before the 1D gapping transition at the edge,
the state has 1 branches of left-movers and $2$ and $1/2$ branches of
right-movers.  After the gapping transition, the same state has $1/2$
branch of left-movers and 2 branches of right-movers.

The phase transition changes the scaling dimension of quasiparticle operators
on the edge, which can in principle be observed in experiment. For FQH edge
states with counterpropagating edge modes it was known that interactions
between the edge branches have to be taken into account to determine the phase
of the edge. It was previously shown that under certain conditions a full
left- and right-moving branch could pair up and open up a gap \cite{KCW9963}.
Here, we showed that half a left- and right-moving branch can pair and become
gapped. This formalism might be generalized further.

\begin{acknowledgments}
We would like to thank C. Chamon for helpful discussions.
This research is partially supported by NSF Grant No.  DMR-0706078.
\end{acknowledgments}

\appendix*

\section{The $U(1)\times SU_2(2)$ edge state}

The edge excitations of the 
$U(1)\times SU_2(2)$ state is described by a charge density  mode $\rho(x)$,
an $S^z$ density mode $\t \rho(x)$,
plus a Majorana fermion $\la(x)$:
\begin{align}
\label{SU2edge}
 [\rho_k,\rho_{k'}]&=\frac{\nu}{2\pi} k\del_{k+k'},\ \ \ \ \nu=\frac12
\nonumber\\ 
 [\t\rho_k,\t\rho_{k'}]&=\frac{1}{2\pi} k\del_{k+k'},
\nonumber\\ 
 \{\la_k,\la_{k'}\}&=\del_{k+k'}, \ \ \ \  \la_k^\dag=\la_{-k}
\nonumber\\ 
 H&=
2\pi \sum_{k>0} [V \rho_{-k}\rho_{k} +\t V \t\rho_{-k}\t\rho_{k})]
+ \sum_{k>0} V_\la k \la_{-k}\la_{k}
\end{align}
There are three electron operators given by
\begin{align*}
 \Psi_{e,3}(x) &=\la(x)e^{2i\phi(x)}  \nonumber\\
 \Psi_{e,1}(x) \pm  i \Psi_{e,2}(x) &=e^{\pm i\t \phi(x)}
e^{2i\phi(x)}  
\nonumber\\
\end{align*}
The $e/4$ quasiparticle operators are given by
\begin{align*}
\psi_{q,1}&=  \si(x) e^{i\frac12 \t \phi(x)} e^{i\frac12 \phi(x)}
\nonumber\\
\psi_{q,2}&=  \si(x) e^{-i\frac12 \t \phi(x)} e^{i\frac12 \phi(x)}
\end{align*}
We find that
\begin{equation*}
 \<\psi^\dag(x,t) \psi(x',t')\>\sim (z-z')^{-g}
\end{equation*}
with
\begin{equation*}
 g=\frac18 + \frac14 +\frac18=\frac12 .
\end{equation*}

%\bibliography{./all,./wencross,./publst,./bibgappedantipf}

\begin{thebibliography}{36}
\expandafter\ifx\csname natexlab\endcsname\relax\def\natexlab#1{#1}\fi
\expandafter\ifx\csname bibnamefont\endcsname\relax
  \def\bibnamefont#1{#1}\fi
\expandafter\ifx\csname bibfnamefont\endcsname\relax
  \def\bibfnamefont#1{#1}\fi
\expandafter\ifx\csname citenamefont\endcsname\relax
  \def\citenamefont#1{#1}\fi
\expandafter\ifx\csname url\endcsname\relax
  \def\url#1{\texttt{#1}}\fi
\expandafter\ifx\csname urlprefix\endcsname\relax\def\urlprefix{URL }\fi
\providecommand{\bibinfo}[2]{#2}
\providecommand{\eprint}[2][]{\url{#2}}

\bibitem[{\citenamefont{Kitaev}(2003)}]{K032}
\bibinfo{author}{\bibfnamefont{A.~Y.} \bibnamefont{Kitaev}},
  \bibinfo{journal}{Ann. Phys. (N.Y.)} \textbf{\bibinfo{volume}{303}},
  \bibinfo{pages}{2} (\bibinfo{year}{2003}).

\bibitem[{\citenamefont{Wen and Niu}(1990)}]{WNtop}
\bibinfo{author}{\bibfnamefont{X.-G.} \bibnamefont{Wen}} \bibnamefont{and}
  \bibinfo{author}{\bibfnamefont{Q.}~\bibnamefont{Niu}},
  \bibinfo{journal}{Phys. Rev. B} \textbf{\bibinfo{volume}{41}},
  \bibinfo{pages}{9377} (\bibinfo{year}{1990}).

\bibitem[{\citenamefont{Nayak et~al.}(2007)\citenamefont{Nayak, Simon, Stern,
  Freedman, and Das Sarma}}]{NSSFD0707.1889}
\bibinfo{author}{\bibfnamefont{C.}~\bibnamefont{Nayak}},
  \bibinfo{author}{\bibfnamefont{S.~H.} \bibnamefont{Simon}},
  \bibinfo{author}{\bibfnamefont{A.}~\bibnamefont{Stern}},
  \bibinfo{author}{\bibfnamefont{M.}~\bibnamefont{Freedman}}, \bibnamefont{and}
  \bibinfo{author}{\bibfnamefont{S.}~ \bibnamefont{Das Sarma}}
  (\bibinfo{year}{2007}), \eprint{arXiv:0707.1889}.

\bibitem[{\citenamefont{Willett et~al.}(1987)\citenamefont{Willett, Eisenstein,
  Str{\"o}rmer, Tsui, Gossard, and English}}]{WES8776}
\bibinfo{author}{\bibfnamefont{R.}~\bibnamefont{Willett}},
  \bibinfo{author}{\bibfnamefont{J.~P.} \bibnamefont{Eisenstein}},
  \bibinfo{author}{\bibfnamefont{H.~L.} \bibnamefont{Str{\"o}rmer}},
  \bibinfo{author}{\bibfnamefont{D.~C.} \bibnamefont{Tsui}},
  \bibinfo{author}{\bibfnamefont{A.~C.} \bibnamefont{Gossard}},
  \bibnamefont{and} \bibinfo{author}{\bibfnamefont{J.~H.}
  \bibnamefont{English}}, \bibinfo{journal}{Phys. Rev. Lett.}
  \textbf{\bibinfo{volume}{59}}, \bibinfo{pages}{1776} (\bibinfo{year}{1987}).

\bibitem[{\citenamefont{Pan et~al.}(2001)\citenamefont{Pan, Yeh, Xia, Stormer,
  Tsui, Adams, Pfeiffer, Baldwin, and West}}]{PYXetal0109}
\bibinfo{author}{\bibfnamefont{W.}~\bibnamefont{Pan}},
  \bibinfo{author}{\bibfnamefont{A.~S.} \bibnamefont{Yeh}},
  \bibinfo{author}{\bibfnamefont{J.~S.} \bibnamefont{Xia}},
  \bibinfo{author}{\bibfnamefont{H.~L.} \bibnamefont{Stormer}},
  \bibinfo{author}{\bibfnamefont{D.~C.} \bibnamefont{Tsui}},
  \bibinfo{author}{\bibfnamefont{E.~D.} \bibnamefont{Adams}},
  \bibinfo{author}{\bibfnamefont{L.~N.} \bibnamefont{Pfeiffer}},
  \bibinfo{author}{\bibfnamefont{K.~W.} \bibnamefont{Baldwin}},
  \bibnamefont{and} \bibinfo{author}{\bibfnamefont{K.~W.} \bibnamefont{West}},
  \bibinfo{journal}{Physica E} \textbf{\bibinfo{volume}{9}}, \bibinfo{pages}{9}
  (\bibinfo{year}{2001}).

\bibitem[{\citenamefont{Moore and Read}(1991)}]{MR9162}
\bibinfo{author}{\bibfnamefont{G.}~\bibnamefont{Moore}} \bibnamefont{and}
  \bibinfo{author}{\bibfnamefont{N.}~\bibnamefont{Read}},
  \bibinfo{journal}{Nucl. Phys. B} \textbf{\bibinfo{volume}{360}},
  \bibinfo{pages}{362} (\bibinfo{year}{1991}).

\bibitem[{\citenamefont{Wen}(1991{\natexlab{a}})}]{Wnab}
\bibinfo{author}{\bibfnamefont{X.-G.} \bibnamefont{Wen}},
  \bibinfo{journal}{Phys. Rev. Lett.} \textbf{\bibinfo{volume}{66}},
  \bibinfo{pages}{802} (\bibinfo{year}{1991}{\natexlab{a}}).

\bibitem[{\citenamefont{Halperin}(1983)}]{H8375}
\bibinfo{author}{\bibfnamefont{B.~I.} \bibnamefont{Halperin}},
  \bibinfo{journal}{Helv. Phys. Acta} \textbf{\bibinfo{volume}{56}},
  \bibinfo{pages}{75} (\bibinfo{year}{1983}).

\bibitem[{\citenamefont{Wen}(2000)}]{Wctpt}
\bibinfo{author}{\bibfnamefont{X.-G.} \bibnamefont{Wen}},
  \bibinfo{journal}{Phys. Rev. Lett.} \textbf{\bibinfo{volume}{84}},
  \bibinfo{pages}{3950} (\bibinfo{year}{2000}).

\bibitem[{\citenamefont{Levin et~al.}(2007)\citenamefont{Levin, Halperin, and
  Rosenow}}]{LHR0706}
\bibinfo{author}{\bibfnamefont{M.}~\bibnamefont{Levin}},
  \bibinfo{author}{\bibfnamefont{B.~I.} \bibnamefont{Halperin}},
  \bibnamefont{and} \bibinfo{author}{\bibfnamefont{B.}~\bibnamefont{Rosenow}},
  \bibinfo{journal}{Phys. Rev. Lett.} \textbf{\bibinfo{volume}{99}},
  \bibinfo{eid}{236806} (\bibinfo{year}{2007}).

\bibitem[{\citenamefont{Lee et~al.}(2007)\citenamefont{Lee, Ryu, Nayak, and
  Fisher}}]{LRNF0707}
\bibinfo{author}{\bibfnamefont{S.-S.} \bibnamefont{Lee}},
  \bibinfo{author}{\bibfnamefont{S.}~\bibnamefont{Ryu}},
  \bibinfo{author}{\bibfnamefont{C.}~\bibnamefont{Nayak}}, \bibnamefont{and}
  \bibinfo{author}{\bibfnamefont{M.~P.~A.} \bibnamefont{Fisher}},
  \bibinfo{journal}{Phys. Rev. Lett.} \textbf{\bibinfo{volume}{99}},
  \bibinfo{eid}{236807} (\bibinfo{year}{2007}).

\bibitem[{\citenamefont{Das Sarma et~al.}(2005)\citenamefont{Sarma, Freedman, and
  Nayak}}]{SFN0502}
\bibinfo{author}{\bibfnamefont{S.}~ \bibnamefont{Das Sarma}},
  \bibinfo{author}{\bibfnamefont{M.}~\bibnamefont{Freedman}}, \bibnamefont{and}
  \bibinfo{author}{\bibfnamefont{C.}~\bibnamefont{Nayak}},
  \bibinfo{journal}{Phys. Rev. Lett.} \textbf{\bibinfo{volume}{94}},
  \bibinfo{pages}{166802} (\bibinfo{year}{2005}).

\bibitem[{\citenamefont{Wen}(1991{\natexlab{b}})}]{Wedgetun}
\bibinfo{author}{\bibfnamefont{X.-G.} \bibnamefont{Wen}},
  \bibinfo{journal}{Phys. Rev. B} \textbf{\bibinfo{volume}{44}},
  \bibinfo{pages}{5708} (\bibinfo{year}{1991}{\natexlab{b}}).

\bibitem[{\citenamefont{Wen}(1992)}]{Wedgerev}
\bibinfo{author}{\bibfnamefont{X.-G.} \bibnamefont{Wen}},
  \bibinfo{journal}{Int. J. Mod. Phys. B} \textbf{\bibinfo{volume}{6}},
  \bibinfo{pages}{1711} (\bibinfo{year}{1992}).

\bibitem[{\citenamefont{de~C.~Chamon et~al.}(1997)\citenamefont{de~C.~Chamon,
  Freed, Kivelson, Sondhi, and Wen}}]{CFK9731}
\bibinfo{author}{\bibfnamefont{C.}~\bibnamefont{de~C.~Chamon}},
  \bibinfo{author}{\bibfnamefont{D.~E.} \bibnamefont{Freed}},
  \bibinfo{author}{\bibfnamefont{S.~A.} \bibnamefont{Kivelson}},
  \bibinfo{author}{\bibfnamefont{S.~L.} \bibnamefont{Sondhi}},
  \bibnamefont{and} \bibinfo{author}{\bibfnamefont{X.-G.} \bibnamefont{Wen}},
  \bibinfo{journal}{Phys. Rev. B} \textbf{\bibinfo{volume}{55}},
  \bibinfo{pages}{2331} (\bibinfo{year}{1997}).

\bibitem[{\citenamefont{Feldman and Kitaev}(2006)}]{FK0603}
\bibinfo{author}{\bibfnamefont{D.~E.} \bibnamefont{Feldman}} \bibnamefont{and}
  \bibinfo{author}{\bibfnamefont{A.}~\bibnamefont{Kitaev}},
  \bibinfo{journal}{Phys. Rev. Lett.} \textbf{\bibinfo{volume}{97}},
  \bibinfo{eid}{186803} (\bibinfo{year}{2006}).

\bibitem[{\citenamefont{Radu et~al.}(2008)\citenamefont{Radu, Miller, Marcus,
  Kastner, Pfeiffer, and West}}]{RMMKPW0803.3530}
\bibinfo{author}{\bibfnamefont{I.~P.} \bibnamefont{Radu}},
  \bibinfo{author}{\bibfnamefont{J.~B.} \bibnamefont{Miller}},
  \bibinfo{author}{\bibfnamefont{C.~M.} \bibnamefont{Marcus}},
  \bibinfo{author}{\bibfnamefont{M.~A.} \bibnamefont{Kastner}},
  \bibinfo{author}{\bibfnamefont{L.~N.} \bibnamefont{Pfeiffer}},
  \bibnamefont{and} \bibinfo{author}{\bibfnamefont{K.~W.} \bibnamefont{West}}
  (\bibinfo{year}{2008}), \eprint{arXiv:0803.3530}.

\bibitem[{\citenamefont{Dolev et~al.}(2008)\citenamefont{Dolev, Heiblum,
  Umansky, Stern, and Mahalu}}]{DHUSM0802.0930}
\bibinfo{author}{\bibfnamefont{M.}~\bibnamefont{Dolev}},
  \bibinfo{author}{\bibfnamefont{M.}~\bibnamefont{Heiblum}},
  \bibinfo{author}{\bibfnamefont{V.}~\bibnamefont{Umansky}},
  \bibinfo{author}{\bibfnamefont{A.}~\bibnamefont{Stern}}, \bibnamefont{and}
  \bibinfo{author}{\bibfnamefont{D.}~\bibnamefont{Mahalu}}
  (\bibinfo{year}{2008}), \eprint{arXiv:0802.0930}.

\bibitem[{\citenamefont{Chamon and Wen}(1994)}]{CW9427}
\bibinfo{author}{\bibfnamefont{C.}~\bibnamefont{Chamon}} \bibnamefont{and}
  \bibinfo{author}{\bibfnamefont{X.-G.} \bibnamefont{Wen}},
  \bibinfo{journal}{Phys. Rev. B} \textbf{\bibinfo{volume}{49}},
  \bibinfo{pages}{8227} (\bibinfo{year}{1994}).

\bibitem[{\citenamefont{Rezayi and Haldane}(2000)}]{RH0085}
\bibinfo{author}{\bibfnamefont{E.~H.} \bibnamefont{Rezayi}} \bibnamefont{and}
  \bibinfo{author}{\bibfnamefont{F.~D.~M.} \bibnamefont{Haldane}},
  \bibinfo{journal}{Phys. Rev. Lett.} \textbf{\bibinfo{volume}{84}},
  \bibinfo{pages}{4685} (\bibinfo{year}{2000}).

\bibitem[{\citenamefont{Morf}(1998)}]{M9805}
\bibinfo{author}{\bibfnamefont{R.~H.} \bibnamefont{Morf}},
  \bibinfo{journal}{Phys. Rev. Lett.} \textbf{\bibinfo{volume}{80}},
  \bibinfo{pages}{1505} (\bibinfo{year}{1998}).

\bibitem[{\citenamefont{Wan et~al.}(2008)\citenamefont{Wan, Hu, Rezayi, and
  Yang}}]{WHRY0816}
\bibinfo{author}{\bibfnamefont{X.}~\bibnamefont{Wan}},
  \bibinfo{author}{\bibfnamefont{Z.-X.} \bibnamefont{Hu}},
  \bibinfo{author}{\bibfnamefont{E.~H.} \bibnamefont{Rezayi}},
  \bibnamefont{and} \bibinfo{author}{\bibfnamefont{K.}~\bibnamefont{Yang}},
  \bibinfo{journal}{Phys.Rev. B} \textbf{\bibinfo{volume}{77}},
  \bibinfo{eid}{165316} (\bibinfo{year}{2008}).

\bibitem[{\citenamefont{chung Kao et~al.}(1999)\citenamefont{Chung Kao, Chang,
  and Wen}}]{KCW9963}
\bibinfo{author}{\bibfnamefont{H.}~\bibnamefont{Chung Kao}},
  \bibinfo{author}{\bibfnamefont{C.-H.} \bibnamefont{Chang}}, \bibnamefont{and}
  \bibinfo{author}{\bibfnamefont{X.-G.} \bibnamefont{Wen}},
  \bibinfo{journal}{Phys. Rev. Lett.} \textbf{\bibinfo{volume}{83}},
  \bibinfo{pages}{5563} (\bibinfo{year}{1999}).

\bibitem[{\citenamefont{Read}(1990)}]{R9002}
\bibinfo{author}{\bibfnamefont{N.}~\bibnamefont{Read}}, \bibinfo{journal}{Phys.
  Rev. Lett.} \textbf{\bibinfo{volume}{65}}, \bibinfo{pages}{1502}
  (\bibinfo{year}{1990}).

\bibitem[{\citenamefont{Fr{\"o}hlich and Zee}(1991)}]{FZ9117}
\bibinfo{author}{\bibfnamefont{J.}~\bibnamefont{Fr{\"o}hlich}}
  \bibnamefont{and} \bibinfo{author}{\bibfnamefont{A.}~\bibnamefont{Zee}},
  \bibinfo{journal}{Nucl. Phys. B} \textbf{\bibinfo{volume}{364}},
  \bibinfo{pages}{517} (\bibinfo{year}{1991}).

\bibitem[{\citenamefont{Fr{\"o}hlich and Kerler}(1991)}]{FK9169}
\bibinfo{author}{\bibfnamefont{J.}~\bibnamefont{Fr{\"o}hlich}}
  \bibnamefont{and} \bibinfo{author}{\bibfnamefont{T.}~\bibnamefont{Kerler}},
  \bibinfo{journal}{Nucl. Phys. B} \textbf{\bibinfo{volume}{354}},
  \bibinfo{pages}{369} (\bibinfo{year}{1991}).

\bibitem[{\citenamefont{Fr{\"o}hlich and Studer}(1993)}]{FS9333}
\bibinfo{author}{\bibfnamefont{J.}~\bibnamefont{Fr{\"o}hlich}}
  \bibnamefont{and} \bibinfo{author}{\bibfnamefont{U.~M.}
  \bibnamefont{Studer}}, \bibinfo{journal}{Rev. of Mod. Phys.}
  \textbf{\bibinfo{volume}{65}}, \bibinfo{pages}{733} (\bibinfo{year}{1993}).

\bibitem[{\citenamefont{Wen}(1995)}]{Wtoprev}
\bibinfo{author}{\bibfnamefont{X.-G.} \bibnamefont{Wen}},
  \bibinfo{journal}{Advances in Physics} \textbf{\bibinfo{volume}{44}},
  \bibinfo{pages}{405} (\bibinfo{year}{1995}).

\bibitem[{\citenamefont{Blok and Wen}(1990{\natexlab{a}})}]{BWkmat1}
\bibinfo{author}{\bibfnamefont{B.}~\bibnamefont{Blok}} \bibnamefont{and}
  \bibinfo{author}{\bibfnamefont{X.-G.} \bibnamefont{Wen}},
  \bibinfo{journal}{Phys. Rev. B} \textbf{\bibinfo{volume}{42}},
  \bibinfo{pages}{8133} (\bibinfo{year}{1990}{\natexlab{a}}).

\bibitem[{\citenamefont{Wen and Zee}(1992)}]{WZclass}
\bibinfo{author}{\bibfnamefont{X.-G.} \bibnamefont{Wen}} \bibnamefont{and}
  \bibinfo{author}{\bibfnamefont{A.}~\bibnamefont{Zee}},
  \bibinfo{journal}{Phys. Rev. B} \textbf{\bibinfo{volume}{46}},
  \bibinfo{pages}{2290} (\bibinfo{year}{1992}).

\bibitem[{\citenamefont{Yang et~al.}(1992)\citenamefont{Yang, Su, and
  Su}}]{YSS9219}
\bibinfo{author}{\bibfnamefont{J.}~\bibnamefont{Yang}},
  \bibinfo{author}{\bibfnamefont{Z.}~\bibnamefont{Su}}, \bibnamefont{and}
  \bibinfo{author}{\bibfnamefont{W.}~\bibnamefont{Su}}, \bibinfo{journal}{Mod.
  Phys. Lett. B} \textbf{\bibinfo{volume}{6}}, \bibinfo{pages}{119}
  (\bibinfo{year}{1992}).

\bibitem[{\citenamefont{Blok and Wen}(1990{\natexlab{b}})}]{BWkmat2}
\bibinfo{author}{\bibfnamefont{B.}~\bibnamefont{Blok}} \bibnamefont{and}
  \bibinfo{author}{\bibfnamefont{X.-G.} \bibnamefont{Wen}},
  \bibinfo{journal}{Phys. Rev. B} \textbf{\bibinfo{volume}{42}},
  \bibinfo{pages}{8145} (\bibinfo{year}{1990}{\natexlab{b}}).

\bibitem[{\citenamefont{Blok and Wen}(1992)}]{BWnab}
\bibinfo{author}{\bibfnamefont{B.}~\bibnamefont{Blok}} \bibnamefont{and}
  \bibinfo{author}{\bibfnamefont{X.-G.} \bibnamefont{Wen}},
  \bibinfo{journal}{Nucl. Phys. B} \textbf{\bibinfo{volume}{374}},
  \bibinfo{pages}{615} (\bibinfo{year}{1992}).

\bibitem[{\citenamefont{Wen}(1999)}]{Wpcon}
\bibinfo{author}{\bibfnamefont{X.-G.} \bibnamefont{Wen}},
  \bibinfo{journal}{Phys. Rev. B} \textbf{\bibinfo{volume}{60}},
  \bibinfo{pages}{8827} (\bibinfo{year}{1999}).

\bibitem[{\citenamefont{Wen et~al.}(1994)\citenamefont{Wen, Wu, and
  Hatsugai}}]{WWHopa}
\bibinfo{author}{\bibfnamefont{X.-G.} \bibnamefont{Wen}},
  \bibinfo{author}{\bibfnamefont{Y.-S.} \bibnamefont{Wu}}, \bibnamefont{and}
  \bibinfo{author}{\bibfnamefont{Y.}~\bibnamefont{Hatsugai}},
  \bibinfo{journal}{Nucl. Phys. B} \textbf{\bibinfo{volume}{422}},
  \bibinfo{pages}{476} (\bibinfo{year}{1994}).

\bibitem[{\citenamefont{Kane and Fisher}(1997)}]{KF9732}
\bibinfo{author}{\bibfnamefont{C.~L.} \bibnamefont{Kane}} \bibnamefont{and}
  \bibinfo{author}{\bibfnamefont{M.~P.~A.} \bibnamefont{Fisher}},
  \bibinfo{journal}{Phys. Rev. B} \textbf{\bibinfo{volume}{55}},
  \bibinfo{pages}{15832} (\bibinfo{year}{1997}).

\end{thebibliography}

%%%%%%%%%%%%%%%%%%%%%%%%%%%%%%%%%%
%%%%%%%%%%%%%%%%%%%%%%%%%%%%%%%%%%
%%%%%%%%%%%%%%%%%%%%%%%%%%%%%%%%%%

\end{document}